# Three-dimensional reconstruction of the Fermi Surface of LaB$_6$

M Monge[1,2], M Biasini[2], G Kontrim-Sznajd[2] and N Sato[3]

[1]*ENEA via don Fiammelli 2 40128 Bologna, Italy*

[2]*Trzebiatowski Institute of Low Temperature and Structure Research P.O.Box 937 Wroclaw, Poland*

[3]*Universidad Carlos III av Universidad 30, Leganes 28911 Spain*

[4]*Physics Department Nagoya University Nagoya 464-8602 Japan*

**More details in**: Materials Science Forum 363-365, pp 582-584 (2001)

The 2-dimensional angular correlations of the positron annihilation radiation (2D-ACAR) on a single crystal of LaB$_6$ was measured for three projections. The 2D-ACAR spectra were subjected to the Van Citter-Gerhard deconvolution alghorithm. From the experimental spectra we produced the 3D $k$-space density $\rho(k)$ via three different methods: i) we reconstructed the 3D electron-positron momentum density $\rho(p)$ via the Cormack method. $\rho(k)$ was then obtained by applying the 3D LCW transformation to $\rho(p)$. ii) The same steps of point i) were repeated adopting a modified Fourier-transform-based algorithm. iii) The 3d $k$-space occupancy $\rho(k)$ was obtained directly by parameterising the Fermi surface volume with prolate ellipsoids whose axes were determined by a least-square fit to the 2D LCW transformations applied to two high symmetry projections. The ellipsoids diameters along high symmetry directions of the cubic Brillouin zone (BZ), as a fraction of the BZ size, were 0.64 (dir. X-M) and 0.82 (dir. Γ-X). The resulting Fermi volume, as a fraction of the BZ volume, was 0.55. These results agree within 1% with those obtained via the de Haas van Alphen experiments.

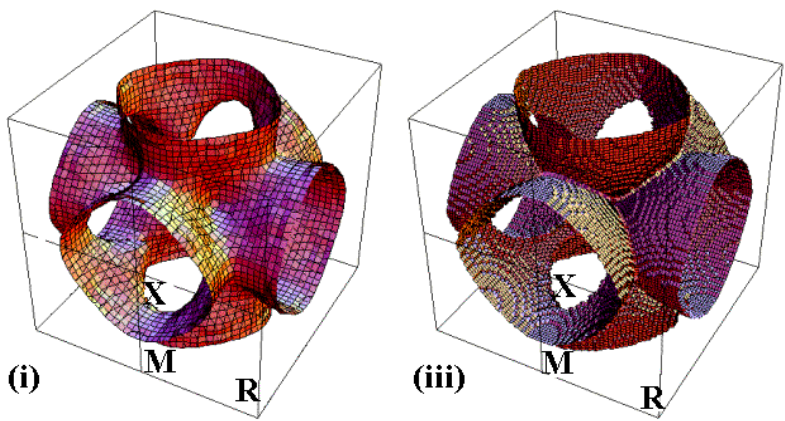

The FS of LaB$_6$ obtained from reconstructed densities (i) and experimental profiles folded into the first Brillouin zone (iii).